\documentclass{ws-ijmpa}

\begin{document}

%------------------------BEGINNING OF CERN TITLE PAGE----------------------

\thispagestyle{empty}

\begin{flushright}
CERN-PH-TH/2005-130\\
hep-ph/0507156
\end{flushright}

\vspace{2.0truecm}
\begin{center}
\boldmath
\large\bf The $B\to\pi\pi,\pi K$ Puzzles: Implications for Hadron 

\vspace*{0.2truecm}

Physics, New Physics and Rare Decays
\unboldmath
\end{center}

\vspace{0.9truecm}
\begin{center}
Robert Fleischer\\[0.1cm]
{\sl CERN, Department of Physics, Theory Unit\\
CH-1211 Geneva 23, Switzerland}
\end{center}

\vspace{0.9truecm}

\begin{center}
{\bf Abstract}
\end{center}

{\small
\vspace{0.2cm}\noindent
The $B$-meson system is an interesting probe for the exploration of strong 
interactions, the quark-flavour sector of the Standard Model, and the search 
for new physics. In this programme, non-leptonic $B$ decays, which are 
particularly challenging from the point of view of QCD, play a key r\^ole. After 
discussing strategies to deal with the corresponding hadronic matrix elements 
of four-quark operators and popular avenues for new physics to manifest itself 
in $B$ decays, we focus on puzzling patterns in the $B$-factory data for 
$B\to\pi\pi,\pi K$ decays; we explore their implications for hadron physics,
new physics and rare $K$ and $B$ decays.
}

\vspace{0.9truecm}

\begin{center}
{\sl Plenary talk at the International Conference on QCD
and Hadronic Physics\\
Peking University, Beijing, China, 16--20 June 2005\\
To appear in the Proceedings (IJMA, World Scientific)}
\end{center}

\vfill
\noindent
CERN-PH-TH/2005-130\\
July 2005

\newpage
\thispagestyle{empty}
\vbox{}
\newpage
 
\setcounter{page}{1}

%------------------------END OF CERN TITLE PAGE------------------------------

\markboth{Robert Fleischer}
{The $B\to\pi\pi,\pi K$ Puzzles: Implications for
Hadron Physics, New Physics and Rare Decays}

\catchline{}{}{}{}{}
%
%%%%%%%%%%%%%%%%%%%%%%%%%%%%%%%%%%%%%%%%%%%%%%%%%%%%%%%%%%%%%%%%%%%%

\boldmath\title{THE $B\to\pi\pi,\pi K$
PUZZLES: IMPLICATIONS FOR HADRON PHYSICS,
NEW PHYSICS AND RARE DECAYS}\unboldmath

\author{\footnotesize R. FLEISCHER}
%\footnote{Typeset names in 8 pt roman, uppercase. Use the footnote to indicate the
%present or permanent address of the author.}

\address{CERN, Department of Physics, Theory Unit\\
CH-1211 Geneva 23, Switzerland}

\maketitle

%\pub{Received (Day Month Year)}{Revised (Day Month Year)}

\begin{abstract}
The $B$-meson system is an interesting probe for the exploration of strong 
interactions, the quark-flavour sector of the Standard Model, and the search 
for new physics. In this programme, non-leptonic $B$ decays, which are 
particularly challenging from the point of view of QCD, play a key r\^ole. After 
discussing strategies to deal with the corresponding hadronic matrix elements 
of four-quark operators and popular avenues for new physics to manifest itself 
in $B$ decays, we focus on puzzling patterns in the $B$-factory data for 
$B\to\pi\pi,\pi K$ decays; we explore their implications for hadron physics,
new physics and rare $K$ and $B$ decays.

\end{abstract}

\boldmath
\section{$B$ Physics in a Nutshell}	
\unboldmath
In this decade, there are continuing huge experimental efforts to explore the
quark-flavour sector of the Standard Model (SM): BaBar (SLAC) and Belle (KEK)
have already collected ${\cal O}(10^8)$ $B\bar B$ pairs, first $B$-physics 
results are coming from run II of the Tevatron (FNAL), second generation
$B$-decay studies will start at the LHC (CERN) in 2007, and there are various
plans to measure rare kaon decays. In this programme, the $B$-meson
system is a particularly interesting playground, with challenging aspects of
strong interactions, avenues to obtain valuable insights into weak interactions,
and probes to search for ``new physics" (NP); for a detailed recent discussion,
see Ref.~\refcite{RF-Lake-Louise}.
A key r\^ole is played by non-leptonic $B$ decays, which may receive 
contributions from tree, QCD penguin and electroweak (EW) penguin 
topologies. In order to deal with these processes, low-energy effective
Hamiltonians are used, which are calculated by means of the operator
product expansion, yielding transition amplitudes of the following 
structure:\cite{BBL-rev}
\begin{equation}\label{LEH}
\langle f|{\cal H}_{\rm eff}|B\rangle=
\frac{G_{\rm F}}{\sqrt{2}}V_{\rm CKM}\sum_{k}C_{k}(\mu)\,
\langle f|Q_{k}(\mu)|B\rangle,
\end{equation}
where $G_{\rm F}$ is Fermi's constant, $V_{\rm CKM}$ a factor containing the 
corresponding elements of the Cabibbo--Kobayashi--Maskawa (CKM) matrix, 
and $\mu$ denotes a renormalization scale. The $Q_k$ are local operators, which 
are generated through the interplay between electroweak interactions and QCD, 
and govern ``effectively'' the decay in question, whereas the Wilson coefficients $C_k(\mu)$ describe the scale-dependent ``couplings" of the interaction vertices 
that are associated with the $Q_k$. In this formalism, the short-distance 
contributions are described by the perturbatively calculable Wilson coefficients
$C_{k}(\mu)$, whereas the long-distance physics arises in the form of 
hadronic matrix elements $\langle f|Q_{k}(\mu)|B\rangle$. These non-perturbative
quantities are the key problem in the theoretical analyses of non-leptonic $B$ 
decays. Although there were interesting recent developments in this field 
through QCD factorization (QCDF),\cite{BBNS} the perturbative 
hard-scattering (PQCD) approach,\cite{PQCD} soft collinear effective theory 
(SCET),\cite{SCET} and QCD light-cone sum-rule methods,\cite{LCSR} as 
discussed at this conference in the plenary talks by Beneke, Cheng, Du and 
Bauer, the $B$-factory data indicate that the theoretical challenge 
remains (see, for instance, Refs.~\refcite{BFRS}--\refcite{CGRS}). 

Fortunately, it is possible to circumvent the calculation of the hadronic
matrix elements for the exploration of CP violation:
\begin{itemize}
\item Amplitude relations can be used to eliminate the 
hadronic matrix elements. We distinguish between exact relations, 
using pure ``tree'' decays  of the kind $B\to KD$ or $B_c\to D_sD$, 
and relations, which follow from the flavour symmetries of strong interactions, 
and involve $B_{(s)}\to\pi\pi,\pi K,KK$ modes. 
\item In the neutral $B_q$ systems ($q\in\{d,s\}$), the interference between 
$B^0_q$--$\bar B^0_q$ mixing and decay processes may lead to
``mixing-induced CP violation''. If a single CKM amplitude dominates the 
decay, the hadronic matrix elements cancel in the corresponding CP asymmeties; 
otherwise we have to use amplitude relations again.
\end{itemize}
These two avenues offer various strategies to ``overconstrain" the unitarity
triangle (UT) of the CKM matrix through studies of CP violation in the $B$ 
system.\cite{RF-Lake-Louise} Moreover, ``rare" decays, which originate 
from loop processes in the SM, provide valuable complementary information;
important examples are $B\to K^\ast \gamma$, $B\to\rho\gamma$, 
$B_{s,d}\to \mu^+\mu^-$ and $K\to\pi\nu\bar\nu$ transitions. 
In the presence of NP effects in the quark-flavour
sector, we expect to encounter discrepancies with the picture emerging from the
CKM mechanism. Popular ways for NP to manifest itself are the following:
\begin{itemize}
\item {\it $B^0_q$--$\bar B^0_q$ mixing:} NP may enter through the exchange
of new particles in box diagrams, which contribute in the SM, or through new
contributions at the tree level, thereby modifying the mixing parameters as follows:
\begin{equation}\label{Dm-Phi-NP}
\Delta M_q=\Delta M_q^{\rm SM}+\Delta M_q^{\rm NP}, \quad
\phi_q=\phi_q^{\rm SM}+\phi_q^{\rm NP}.
\end{equation}
Whereas the NP contribution $\Delta M_q^{\rm NP}$ to the mass difference
would affect the determination of one UT side, the NP contribution 
$\phi_q^{\rm NP}$ to the weak mixing
phase would enter the mixing-induced CP asymmetries. Because of the remarkable agreement between the direct determination of the UT angle $\beta$ through
$B_d\to J/\psi K_{\rm S}$ and the CKM fits,\cite{fits} the space for NP is getting 
smaller and smaller in the $B_d$ system. On the other hand, the $B_s$ system is still
essentially unexplored, and will be a key target for LHCb.
\item {\it Decay amplitudes:} NP has typically a small effect if SM tree processes
play the dominant r\^ole, as in $B_d\to J/\psi K_{\rm S}$. On the other hand, 
we encounter potentially large effects in the flavour-changing neutral-current (FCNC) 
sector. For instance, new particles may enter in penguin diagrams, or new FCNC 
processes may arise at the tree level. Interestingly, there are hints in the current
$B$-factory data for such effects. In particular, Belle results for $B_d\to \phi K_{\rm S}$
raise the question of whether 
$(\sin2\beta)_{\phi K_{\rm S}}=(\sin2\beta)_{\psi K_{\rm S}}$, 
and the branching ratios of certain $B\to\pi K$ decays show a puzzling pattern.
\end{itemize}
In the following, we will focus on the latter ``$B\to\pi K$ puzzle", which was 
already indicated by the first CLEO data for the $B^0_d\to \pi^0 K^0$ decay
in 2000,\cite{BF00} and received a lot of attention recently (see, for instance, 
Refs.~\refcite{BeNe}--\refcite{BCLL}).

\begin{figure}[h] 
     \centering
     \vspace*{0.1truecm}
   \includegraphics[width=10.0truecm]{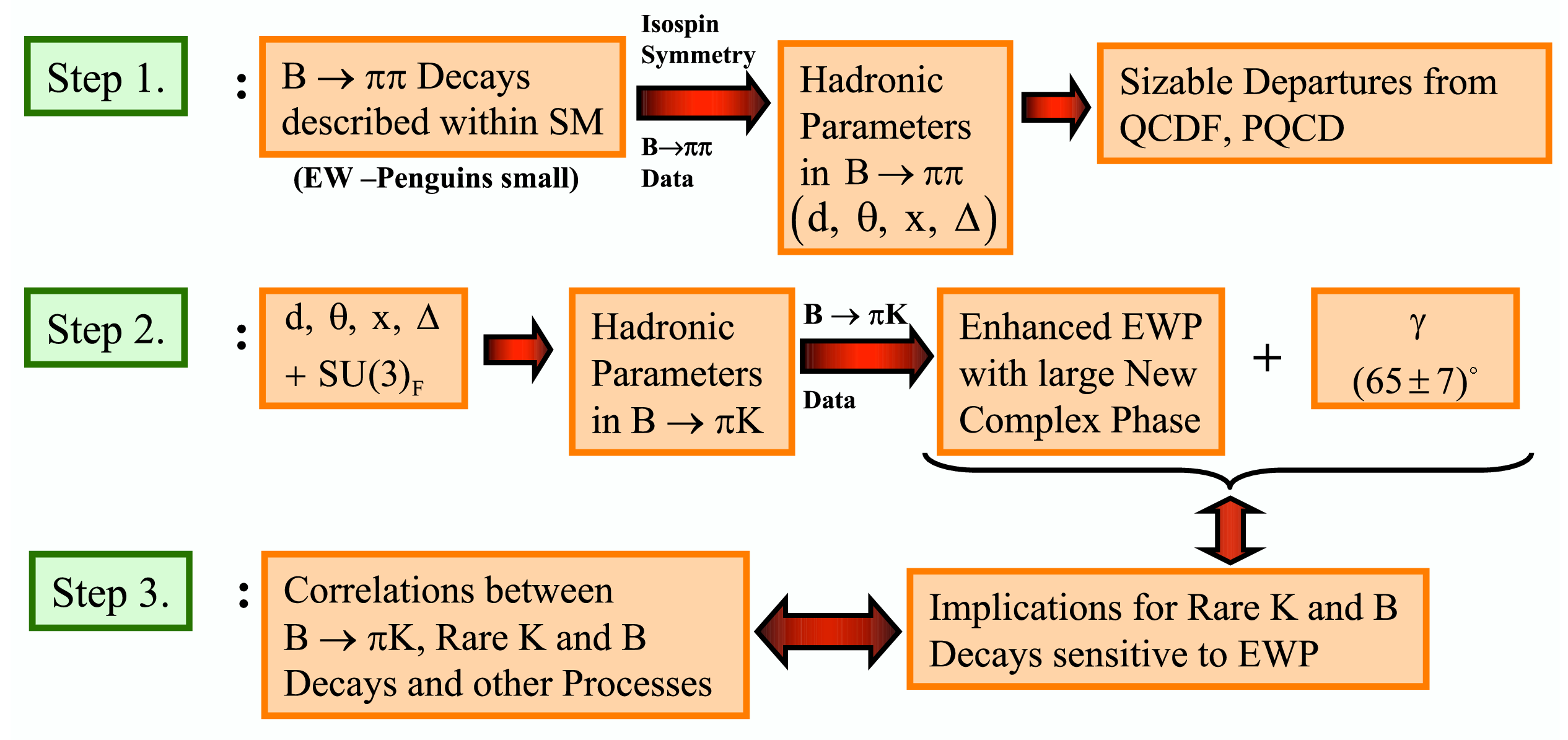} 
    \caption{The structure of a strategy to analyse the 
    $B\to\pi K$ puzzle, as discussed in the text.}\label{fig:chart}
\end{figure}

\boldmath
\section{A Strategy for the Exploration of the $B\to\pi K$ Puzzle}
\unboldmath
In order to analyse the puzzling patterns in the $B\to\pi K$ data, we use
the strategy in three subsequent steps developed in Ref.~\refcite{BFRS}, as
illustrated in Fig.~\ref{fig:chart}; the numerical values refer to the recent update 
given in Ref.~\refcite{BFRS-update}.
\subsection{Step 1: $B\to\pi\pi$}\label{ssec:Bpipi-puzzle}
The $B\to\pi\pi$ system offers three decay channels, 
$B^+\to\pi^+\pi^0$, $B^0_d\to\pi^+\pi^-$ and $B^0_d\to\pi^0\pi^0$,
as well as their CP conjugates. Consequently, we may introduce the following
two independent ratios of the corresponding CP-averaged branching ratios:
\begin{equation}\label{R-pipi}
R_{+-}^{\pi\pi}\equiv
2\left[\frac{\mbox{BR}(B^\pm\to\pi^\pm\pi^0)}{\mbox{BR}
(B_d\to\pi^+\pi^-)}\right]\frac{\tau_{B^0_d}}{\tau_{B^+}}, \quad
R_{00}^{\pi\pi}\equiv2\left[\frac{\mbox{BR}(B_d\to\pi^0\pi^0)}{\mbox{BR}
(B_d\to\pi^+\pi^-)}\right].
\end{equation}
The branching ratios for $B_d\to\pi^+\pi^-$ and $B_d\to\pi^0\pi^0$ are found 
to be surprisingly small and large, respectively, whereas the one for
$B^\pm\to\pi^\pm\pi^0$ is in accordance with theoretical estimates. This feature
is the ``$B\to\pi\pi$ puzzle". In addition to the CP-conserving observables in 
(\ref{R-pipi}), we may also exploit the following CP-violating 
observables of the $B_d\to\pi^+\pi^-$ decay:
\begin{equation}\label{ACP-pipi}
\frac{\Gamma(B^0_d(t)\to \pi^+\pi^-)-
\Gamma(\bar B^0_d(t)\to \pi^+\pi^-)}{\Gamma(B^0_d(t)\to \pi^+\pi^-)+
\Gamma(\bar B^0_q(t)\to \pi^+\pi^-)}
={\cal A}_{\rm CP}^{\rm dir}\,\cos(\Delta M_d t)+{\cal A}_{\rm CP}^{\rm 
mix}\,\sin(\Delta M_d t).
\end{equation}
The experimental picture of these CP asymmetries is not yet fully settled.\cite{HFAG}
However, their theoretical interpretation discussed below yields constraints 
for the UT, in excellent agreement with the CKM fits obtained within the SM. 

Using the isospin flavour symmetry of strong interactions, the observables in
(\ref{R-pipi}) and (\ref{ACP-pipi}) depend on two (complex) hadronic parameters,
$de^{i\theta}$ and $xe^{i\Delta}$, which describe --  sloppily speaking -- the 
ratio of penguin to colour-allowed tree amplitudes and the ratio of  
colour-suppressed to colour-allowed tree amplitudes, respectively. It is possible 
to extract these quantities {\it cleanly} and {\it unambiguously} from the data:
\begin{equation}\label{Bpipi-par-det}
d=0.51^{+0.26}_{-0.20},\quad \theta=+(140^{+14}_{-18})^\circ,\quad x=
1.15^{+0.18}_{-0.16},\quad \Delta=-(59^{+19}_{-26})^\circ;
\end{equation}
a similar picture is also found by other authors.\cite{ALP}$^{\mbox{--}}$\cite{CGRS} 
In particular the impressive strong phases give an unambiguous signal for large deviations from ``factorization". In recent QCDF\cite{busa} and 
PQCD\cite{kesa} analyses, the following numbers were obtained:
\begin{equation}
\left.d\right|_{\rm QCDF}=0.29\pm0.09, \quad
\left.\theta\right|_{\rm QCDF}=-\left(171.4\pm14.3\right)^\circ, 
\end{equation}
\begin{equation}
\left.d\right|_{\rm PQCD}=0.23^{+0.07}_{-0.05}, \quad
+139^\circ < \left.\theta\right|_{\rm PQCD} < +148^\circ,
\end{equation}
which depart significantly from the experimental pattern in (\ref{Bpipi-par-det}).

Having the hadronic parameters given in (\ref{Bpipi-par-det}) at hand, the CP-violating
asymmetries of the $B_d\to\pi^0\pi^0$ channel can be predicted:
\begin{equation}\label{Adir00-pred}
\left.{\cal A}_{\rm CP}^{\rm dir}(B_d\to\pi^0\pi^0)
\right|_{\rm SM}=-0.28^{+0.37}_{-0.21}, \quad
\left.{\cal A}_{\rm CP}^{\rm mix}(B_d\to\pi^0\pi^0)\right|_{\rm SM}=
-0.63^{+0.45}_{-0.41},
\end{equation}
offering the exciting perspective of large CP violation in this decay. The first results 
for the direct CP asymmetry were recently reported by the BaBar and Belle
collaborations,\cite{Bpi0pi0-exp}  corresponding to the average of 
${\cal A}_{\rm CP}^{\rm dir}(B_d\to \pi^0\pi^0)=-(0.28\pm0.39)$,\cite{HFAG} 
which is in encouraging agreement with (\ref{Adir00-pred}). In the future, more 
accurate input data will allow us to make much more stringent predictions.

\subsection{Step 2: $B\to\pi K$}\label{ssec:BpiK-puzzle}
In contrast to the $B\to\pi\pi$ modes, which originate from $b\to d$ processes,
we have to deal with $b\to s$ transitions in the case of the $B\to\pi K$ system. 
Consequently, these decay classes differ in their CKM structure and exhibit a 
different dynamics. In particular, the $B\to\pi K$ decays are dominated by 
QCD penguins. Concerning the EW penguins, we distinguish between the following
cases: in $B^0_d\to\pi^-K^+$, $B^+\to\pi^+K^0$ transitions, the EW penguin 
amplitudes are colour-suppressed and are hence expected to be tiny. 
On the other hand, EW penguins contribute also in colour-allowed form to the
$B^+\to\pi^0K^+$, $B^0_d\to\pi^0K^0$ system and have therefore a significant 
impact on these modes.

The starting point of our $B\to\pi K$ analysis is given by the hadronic 
parameters determined in Subsection~\ref{ssec:Bpipi-puzzle} and the
CKM fits of the UT, which are only insignificantly affected by EW penguins. 
We then use the following working hypotheses:
\begin{itemize}
\item[(i)] $SU(3)$ flavour symmetry of strong interactions;
\item[(ii)] neglect of penguin annihilation and exchange topologies.
\end{itemize}
Internal consistency checks of these assumptions can be performed. They
are nicely satisfied by the current data, and do not indicate any anomalous 
behaviour. We may then determine the hadronic $B\to\pi K$ parameters 
through their $B\to\pi\pi$ counterparts, allowing us to predict 
the $B\to\pi K$ observables in the SM. 

In the case of the observables with a tiny impact of EW penguins, we obtain
agreement between the SM predictions and the data. In particular,  our prediction 
${\cal A}_{\rm CP}^{\rm dir}(B_d\to\pi^\mp K^\pm)=+0.127^{+0.102}_{-0.066}$ 
agrees nicely with the measurements of this direct CP asymmetry, which
was observed by BaBar and Belle in the summer of 2004,\cite{CP-dir-B}  
with the average value ${\cal A}_{\rm CP}^{\rm dir}(B_d\to\pi^\mp K^\pm)=
+0.113\pm 0.01$. Moreover, assumptions (i) and (ii) listed above imply 
\begin{equation}
H\propto\underbrace{\left(\frac{f_K}{f_\pi}\right)^2\left[\frac{\mbox{BR}
(B_d\to\pi^+\pi^-)}{\mbox{BR}(B_d\to\pi^\mp K^\pm)}
\right]}_{\mbox{0.38$\pm$0.04}}=
\underbrace{-\left[\frac{{A}_{\rm CP}^{\rm dir}(B_d\to\pi^\mp 
K^\pm)}{{A}_{\rm CP}^{\rm dir}(B_d\to\pi^+\pi^-)}
\right]}_{\mbox{0.31$\pm$0.11}}.
\end{equation}
The experimental numbers indicated in this expression give us further confidence 
in our working hypothesis. Furthermore, $H$ allows us to convert the 
$B_d\to\pi^+\pi^-$ CP asymmetries into a value of $\gamma$, 
in excellent agreement with the fits for the UT. On the other hand, a moderate numerical 
discrepancy arises for the ratio $R$ of the CP-averaged $B_d\to\pi^\mp K^\pm$, $B^\pm\to\pi^\pm K$ branching ratios. This feature suggests a sizeable impact of a
hadronic parameter, $\rho_{\rm c}e^{i\theta_{\rm c}}$, which enters the 
most general parametrization of the $B^+\to\pi^+K^0$ amplitude. It can be 
constrained through the direct CP asymmetry of the decay $B^\pm\to\pi^\pm K$ 
and the emerging $B^\pm\to K^\pm K$ signal, and actually shifts the predicted 
value of $R$ towards the data. 

Let us now turn to those observables that are significantly affected by
EW penguins. The key quantities are the following ratios:\cite{BF98}
\begin{equation}
R_{\rm c}\equiv2\left[\frac{\mbox{BR}(B^+\to\pi^0K^+)+
\mbox{BR}(B^-\to\pi^0K^-)}{\mbox{BR}(B^+\to\pi^+ K^0)+
\mbox{BR}(B^-\to\pi^- \bar K^0)}\right]\stackrel{\rm Exp}{=}1.00\pm0.08
\end{equation}
\begin{equation}
R_{\rm n}\equiv\frac{1}{2}\left[\frac{\mbox{BR}(B_d^0\to\pi^- K^+)+
\mbox{BR}(\bar B_d^0\to\pi^+ K^-)}{\mbox{BR}(B_d^0\to\pi^0K^0)+
\mbox{BR}(\bar B_d^0\to\pi^0\bar K^0)}\right]\stackrel{\rm Exp}{=}0.79\pm0.08,
\end{equation}
where the EW penguin contributions enter in colour-allowed form
through the decays with $\pi^0$ mesons in the final states. Theoretically,
the EW penguin effects are described by the following parameters:
\begin{equation}
q\stackrel{\rm SM}{=}0.69, \quad
\phi\stackrel{\rm SM}{=}0^\circ,
\end{equation}
where $q$, which can be calculated in the SM with the help of the $SU(3)$ 
flavour symmetry,\cite{NR} measures the ``strength'' of the EW penguins 
with respect to the tree contributions, and $\phi$ is a CP-violating weak phase 
with an origin lying beyond the SM. EW penguin topologies offer an interesting 
avenue for NP to manifest itself, as has already been known for 
several years.\cite{FM-BpiK-NP,trojan}

\begin{figure}[ht] 
   \centering
    \includegraphics[width=8.2truecm]{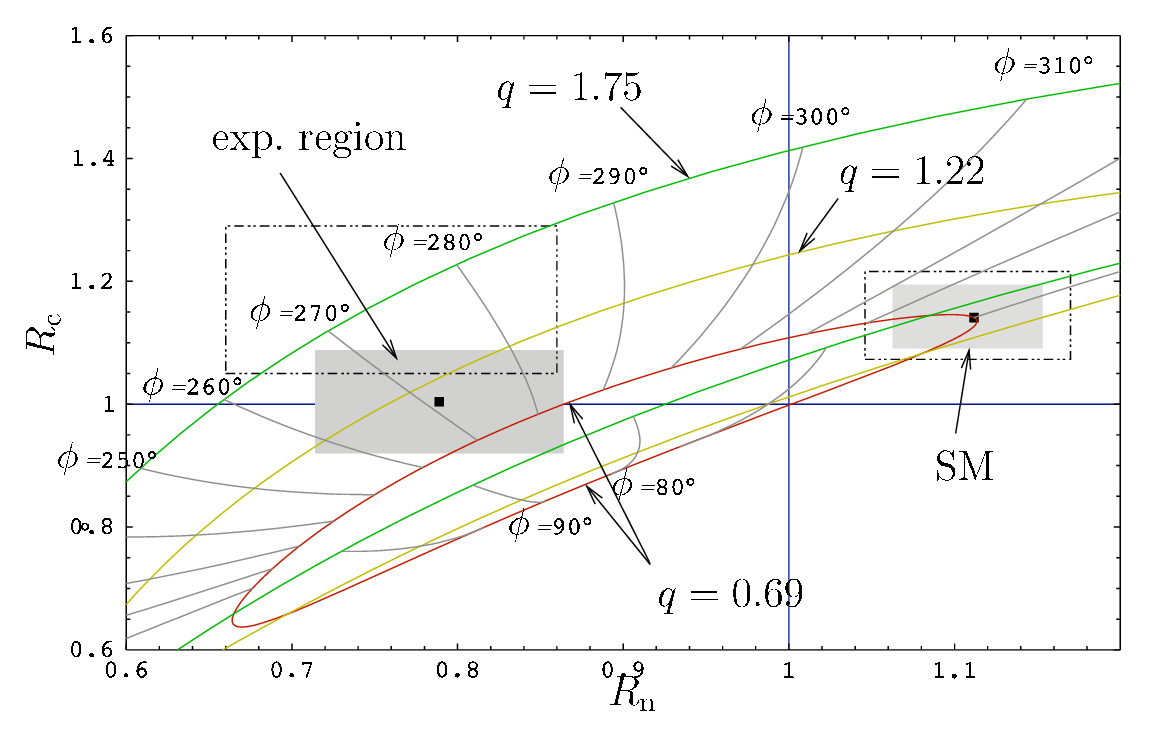} 
   \vspace*{-0.2truecm}
   \caption{The situation in the $R_{\rm n}$--$R_{\rm c}$ plane,
   as discussed in the text.}
   \label{fig:RnRc-update}
\end{figure}

In Fig.~\ref{fig:RnRc-update}, we show the current situation in the 
$R_{\rm n}$--$R_{\rm c}$ plane: the experimental ranges and those 
predicted in the SM are indicated in grey,\cite{BFRS-update} and the dashed 
lines serve as a reminder of the corresponding ranges in Ref.~\refcite{BFRS}; 
the central values for the SM prediction have hardly moved, while their
uncertainties have been reduced a bit. Moreover, we show contours 
for values of $q=0.69$, $q=1.22$ and $q=1.75$, with
$\phi \in [0^\circ,360^\circ]$. We observe that we arrive no longer
at a nice agreement between our SM predictions and the experimental
values. However, as becomes obvious from the contours
in Fig.~\ref{fig:RnRc-update}, this discrepancy can be resolved if we allow for
NP in the EW penguin sector, i.e.\ if we keep $q$ and $\phi$ as free
parameters. Following these lines, the successful picture described above 
would not be disturbed, and we obtain full agreement between the theoretical 
values of $R_{\rm n, c}$ and the data. The corresponding values of $q$
and $\phi$ are given as follows:
\begin{equation}\label{q-phi-det}
q = 1.08\,^{+0.81}_{-0.73}, 
\quad 
\phi = -(88.8^{+13.7}_{-19.0})^\circ,
\end{equation}
where in particular the large CP-violating phase would be a
striking signal of NP. These parameters then allow us to predict 
also the CP-violating observables of the $B^\pm\to\pi^0 K^\pm$
and $B_d\to\pi^0K_{\rm S}$ decays,\cite{BFRS-update} which should
provide useful future tests of this scenario. Particularly promising in this respect 
are rare $K$ and $B$ decays.

\subsection{Step 3: Rare $K$ and $B$ Decays}\label{ssec:rareKB}
In order to explore the implications for rare $K$ and $B$ decays, we
assume that NP enters the EW penguin sector through enhanced 
$Z^0$ penguins with a new CP-violating phase. This scenario was already
considered in the literature, where model-independent analyses and 
studies within SUSY were presented.\cite{Z-pen-analyses,BuHi} 
In our strategy, we determine the short-distance function $C$ characterizing 
the $Z^0$ penguins through the $B\to\pi K$ data. Performing a 
renormalization-group analysis,\cite{BFRS-I} we obtain
\begin{equation}\label{RG}
C(\bar q)= 2.35~ \bar q e^{i\phi} -0.82 \quad\mbox{with}\quad 
\bar q= q \left[\frac{|V_{ub}/V_{cb}|}{0.086}\right].
\end{equation}
If we then evaluate the relevant box-diagram contributions within the SM 
and use (\ref{RG}), we can calculate the short-distance functions:
\begin{equation}\label{X-C-rel}
X=2.35~ \bar q e^{i\phi} -0.09 \quad \mbox{and} \quad 
Y=2.35~ \bar q e^{i\phi} -0.64,
\end{equation}
which govern the rare $K$, $B$ decays with $\nu\bar\nu$ and $\ell^+\ell^-$ 
in the final states, respectively. In the SM, we have $C=0.79$, $X=1.53$
and $Y=0.98$, with {\it vanishing} CP-violating phases. 
If we impose constraints from the data for rare decays, in particular those on
$|Y|$ following from $B\to X_s\mu^+\mu^-$, the following picture arises:
\begin{equation}\label{q-phi-RD}
\bar q= 0.92^{+0.07}_{-0.05},\quad
\phi=-(85^{+11}_{-14})^\circ.
\end{equation}
The corresponding predictions of $R_{\rm c}$ and $R_{\rm n}$ agree nicely
with the data reported at the ICHEP '04 conference.\cite{ICHEP04}  
We may encounter significant deviations from the SM expectations for 
certain rare decays, with a pattern characteristic of our NP scenario. 
The most spectacular effects are the following ones:\cite{BFRS,BFRS-update,ISU,CCG}
\begin{itemize}
\item $K\to\pi\nu\bar\nu$: enhancement of  $\mbox{BR}(K_{\rm L}\to\pi^0\nu\bar\nu)$
by up to one order of magnitude, and strong violation of the 
relation $(\sin2\beta)_{\pi\nu\bar\nu}=(\sin2\beta)_{\psi K_{\rm S}}$.
\item $K_{\rm L}\to\pi^0e^+e^-$ and $K_{\rm L}\to\pi^0\mu^+\mu^-$: now governed
by direct CP violation, with branching ratios enhanced by ${\cal O}(3)$. 
\item $B_d\to K^\ast\mu^+\mu^-$: a forward--backward CP asymmetry
can be very large.
\item $B\to X_{s,d}\nu\bar\nu$ and $B_{s,d}\to \mu^+\mu^-$:
branching ratios may be enhanced by factors as large as ${\cal O}(2)$ and 
${\cal O}(5)$, respectively. 
\end{itemize}

\section{Conclusions}
The $B\to\pi\pi$ data allow the clean extraction of hadronic
parameters, which indicate large non-factorizable effects -- thereby resolving the
``$B\to\pi\pi$ puzzle" -- and favour large CP violation in $B_d\to\pi^0\pi^0$. The 
resulting SM analysis of the $B\to\pi K$ system agrees with the current data, 
with the {\it exception} of the observables receiving sizeable EW penguin 
contributions, which is another manifestation of the ``$B\to\pi K$ puzzle". It can 
be resolved through NP in the EW penguin sector involving a large CP-violating 
phase. If NP enters through $Z^0$ penguins, we expect significant NP effects in 
rare $K$ and $B$ decays, with a pattern that is characteristic of this scenario. 
It will be interesting to confront these results with future, more accurate data.

\section*{Acknowledgements}
I would like to thank the organizers for inviting me to this most interesting
conference. I am particularly grateful to Cai-Dian Lu for his great help in
arranging my stay in China, and to Junxiao Chen, Ying Li, Yue-Long Shen
and Xian-Qiao Yu for their great efforts to give me such an excellent
impression of Beijing.

\end{document}